 \documentclass{article}
 \usepackage{emulateapjnew,apjfonts,rotating,epsf}

\newcommand{\sig}{\:\lower0.6ex\hbox{$\stackrel{\textstyle >}{\sim}$}\:}
\newcommand{\sil}{\:\lower0.6ex\hbox{$\stackrel{\textstyle <}{\sim}$}\:}
\newcommand{\sigs}{\:\lower0.4ex\hbox{$\stackrel{\scriptstyle
      >}{\scriptstyle \sim}$}\,}
\newcommand{\sils}{\:\lower0.4ex\hbox{$\stackrel{\scriptstyle
      <}{\scriptstyle \sim}$}\,}

\begin{document}

\title{ Are dwarf spheroidal  galaxies dark matter dominated or
remnants of disrupted larger satellite galaxies? -- A possible test.} 

\author{Ralf S.\ Klessen\altaffilmark{1,2} and HongSheng Zhao\altaffilmark{3}}

\altaffiltext{1}{UCO/Lick Observatory, University of California at Santa Cruz, 
 CA 95064, U.S.A. ({\tt ralf@ucolick.org})} 
\altaffiltext{2}{Max-Planck-Institut f{\"u}r
 Astronomie, K{\"o}nigstuhl 17, 69117 Heidelberg, Germany}
\altaffiltext{3}{Institute of Astronomy, Cambridge, CB3 0HA, UK ({\tt
 hsz@ast.cam.ac.uk})}

\begin{abstract}
  The failure of standard cosmolocical models in accounting for the
  statistics of dwarf satellites and the rotation curve of gas-rich
  dwarf galaxies in detail has led us to examine whether earlier
  non-equilibrium models of dwarf spheroidal satellites without any
  dark matter should be reconsidered in more detail.  Such models can
  explain the high dispersion of the dwarf spheroids by the projection
  of  disrupted tidal debris.  We show in the case of Milky Way
  satellites, that these models predict a significant spread in the
  apparent magnitude of horizontal branch stars which is correlated
  with sky position and velocity. In particular, the models produce a
  strong correlation of radial velocity with the long axis of the
  dwarf.  Current data do not set strong enough constraint on models,
  but we suggest that photometric and spectroscopic surveys of
  extra-tidal stars of nearby dwarf spheroids in the Milky Way and
  Andromeda can falsify these models without dark matter.
\end{abstract}


\section{Introduction}
\label{sec:intro}

  At least nine dwarf spheroidal galaxies (dSph's) are known to orbit
  the Milky Way at distances ranging from a few tens to a few hundred
  kpc. Their velocity dispersions and stellar masses are similar to
  those seen in globular clusters, however, they are about two orders
  of magnitude more extended (e.g.\ Irwin \& Hatzidimitriou 1995,
  Grebel 1997, Mateo 1998).  Under the standard assumption of virial
  equilibrium the derived mass-to-light ratios are extremely high (up
  to $M/L \approx 100$) implying the presence of huge quantities of
  dark matter.  However, the assumption of virial equilibrium may not
  be valid for satellite galaxies: they could be significantly
  perturbed by Galactic tides or even not be bound at all, but instead
  be the remnants of disrupted larger satellite galaxies.  Indeed, at
  least one dSph (Sagittarius, see Ibata, Gilmore, \& Irwin 1994) is
  known to be torn apart by tidal interaction with the Milky Way and
  others appear heavily distorted (e.g.\ Ursa Minor,
  Mart{\'\i}nez-Delgado et al.\ 2001a).
  
  The non-equilibrium models have received little attention partly
  because there is plenty of evidence for dark matter in halos of
  high-brightness galaxies from the study of, e.g., gravitational
  lensing.  It appears unnecessary to explain the missing dynamical
  matter in dwarfs with a different theory.  However, recent studies
  show that it is problematic to introduce dark matter in dwarfs
  (e.g.\ van den Bosch et al.\ 2000, McGaugh \& de\ Blok 2001), and in
  fact many have tried to remove the undesirable cusp in cold dark
  matter models by including feedback from baryonic material and
  from massive black holes or introducing modifications to the
  gravitational force law (Mateo 1998).  It is thus interesting to
  reexamine options without dark matter to start with in dwarf
  galaxies.
  
  The `tidal scenario' has been studied by a variety of authors (e.g.\ 
  Kuhn 1993, Oh, Lin \& Aarseth 1995, Piatek \& Pryor 1995, Helmi \&
  White 1999, Johnston, Sigursson, \& Hernquist 1999, Bekki, Cough, \&
  Drinkwater 2001).  In particular Kroupa (1997, hereafter K97) and
  Klessen \& Kroupa (1998, hereafter KK98) have followed the long-term
  evolution of low-mass satellite galaxies in the Galactic tidal field
  in a series of high-resolution numerical simulations.  A remnant
  containing a few percent of the initial mass prevails as a
  long-lived and distinguishable entity for a period of several
  billion years after the initial satellite dissolves.  They proposed
  that what appears to be a bound dSph galaxy to an terrestrial
  observer may in fact be this elongated tidal remnant.  This model
  successfully demonstrates that high velocity dispersions in dSphs
  may be obtained {\it without any} internal dark matter.  And it
  accounts for the distorted morphology of e.g., the double-peaked
  Ursa Minor, as well (Kleyna et al. 1998).
  It is surprising such an extreme model passes
  the known observational test so well.  
  It is the aim of this paper to give a first discussion of additional
  tests to falsify/confirm such models.

\begin{figure*}[t]
\unitlength1.0cm
\begin{center}
\begin{picture}(16.0,7.9)
\put(-3.0, -16.0){\epsfbox{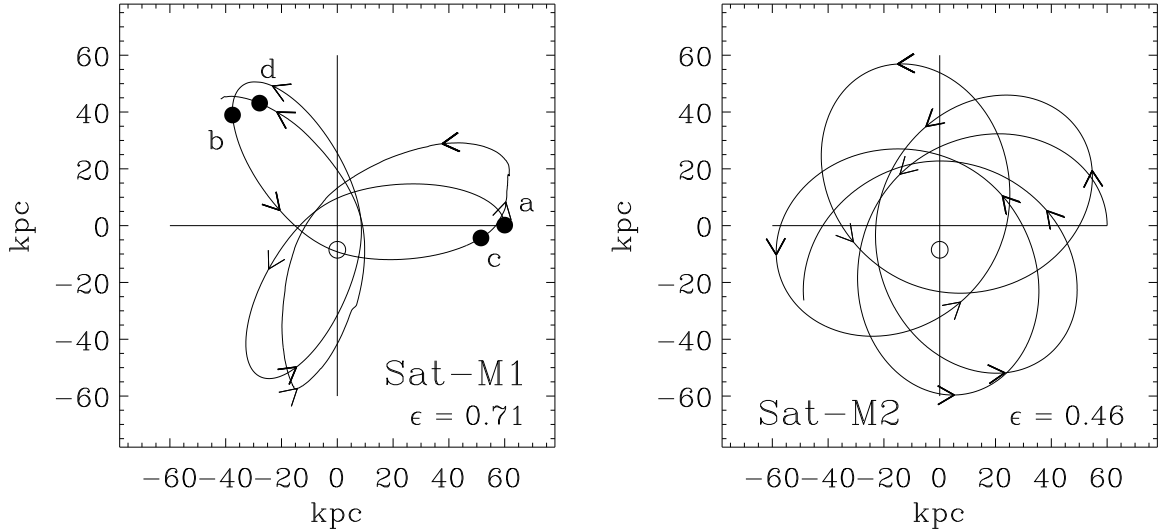}}
\end{picture}
\end{center}
\caption{\label{fig:trajectories}Center-of-density trajectories of the two dSph galaxies
  {\em Sat-M1} and {\em Sat-M2}.  The arrows indicate indicate time
  intervals of 500$\,$Gyr. For {\em Sat-M1}, the  four
  snapshots shown in Figures \ref{fig:contours}, \ref{fig:CMD}, and
  \ref{fig:velocity} are indicated by filled dots. The position of the
  "observer on Earth" is indicate by a open circle.  }
\end{figure*}

\section{Numerical model}
\label{sec:model}

We follow the evolution and disintegration of a satellite dwarf galaxy
in the tidal field of our Galaxy. The satellite has initially a mass
of $10^7\,$M$_{\odot}$, and a Plummer law density profile with core
radius $R_{\rm c} = 0.3\,$kpc and truncation radius $R_{\rm t} =
1.5\,$kpc. The central velocity dispersion in equilibrium is
$\sigma_{\rm c} = 4.4\,$km$\,$s$^{-1}$. The stars have a mass-to-light
ratio $M/L = 3$ and there is {\em no} dark matter component in the dSph.  
The Milky Way, however, does contain dark matter. It dominates the
gravitational potential at large scales.  We describe it as isothermal
sphere with circular velocity of $220\,$km$\,$s$^{-1}$, and adopt a
core radius of 5$\,$pc to obtain a total halo mass of $2.85 \times
10^{12}\,$M$_{\odot}$ within a distance of 250$\,$kpc.

The satellite galaxy consists of 131$\,$072 particles. The equation of
motion is solved by direct summation using the special-purpose
hardware device GRAPE-3 (Sugimoto et al.\ 1990, Ebisuzaki et al.\ 
1993) under the influence of a external rigid Galactic potential. For
further details see KK98.

Here we focus on two model galaxies. {\em Sat-M1} is on a highly
eccentric orbit with $\epsilon = 0.71$ and {\em Sat-M2} has a more
circular trajectory with $\epsilon = 0.46$ (Table 1 in KK98). We
follow their evolution for many orbital periods well into the phase
when the satellites are completely dissolved by tidal shocking.  The
dwarf galaxies are initially placed at a galactocentric distance of
60$\,$kpc with tangential velocities of $60\,$km$\,$s$^{-1}$ and
$100\,$km$\,$s$^{-1}$, respectively (see Table 1 in KK98).  The
trajectories of the stellar density maximum are depicted in Figure
\ref{fig:trajectories}.  The position of a hypothetical observer on
``Earth'' is indicated by the open circle.

\section{Synthetic CMD's}
\label{sec:synthetic-CMD}
Studying the effect of distance variations on the distribution of
stars in the color-magnitude diagram (CMD) requires a template CMD
that is well defined and free of additional bias. Instead of
constructing CMD's from population synthesis models, we found it more
appropriate to use data of a well observed existing astrophysical
object and chose the Galactic globular cluster M3 (i.e.~NGC$\,$5272).
Globular clusters are massive but compact in size and their stars are
more or less coeval. Hence, their CMD's are well populated and not
influenced by variations in the stellar distance moduli or by a
complicated star formation history. M3 is particularly well suited. It
is a classical prototype of an old (Pop II) cluster. It is a well
observed object with very accurate photometric data available.
Bounanno et al.~(1994) reanalyzed the Mt.~Palomar and Mt.~Wilson
original plates (Sandage 1953) to obtain a sample of more than
$10\,000$ M3 stars with $V>21.5$. We use their data to construct our
templated CMD.

M3 has a very {\em thin} and well defined HB with a large range in
colors, $-0.4 \sils (B-V) \sils 0.6$ (see also Sandage \& Katem
1982). The faint end of the HB lies at $V \approx 17.5$ with
photometric errors of typically $\sils 0.02$ mag. The average
``thickness'' of the HB is $\Delta V \approx 0.08$, defined as the
standard deviation from the mean value.  The distance modulus of the
cluster is $(m - M)_0 = 14.94$. Age and metallicity are estimated as
$t=18.7 \pm 3.5\,$Gyr and $\langle [{\rm Fe/H}]\rangle = -1.66 \pm
0.10$.  M3 falls therefore right into the range relevant for the
Galactic dSph's.

To construct {\em synthetic} CMD's from the numerical models of dSph
galaxies without dark matter, we use the CMD of M3 as starting point and
select all stars with $V\le18$. Then we take the model galaxy, project
it onto the sky of an hypothetical observer on ``Earth''. This
procedure is introduced in K97 and KK98. The synthetic CMD
of the dSph galaxy finally is obtained by convolving its stars with the
template CMD using the appropriate distance moduli, i.e.\ for each
star in the model galaxy we randomly select one star in the template
CMD and modify its apparent luminosity according to its distance to
the observer.

\begin{figure*}[t]
\unitlength1cm
\begin{center}
\begin{picture}(17.0,15.0)
\put( 1.0, 0.0){\epsfbox{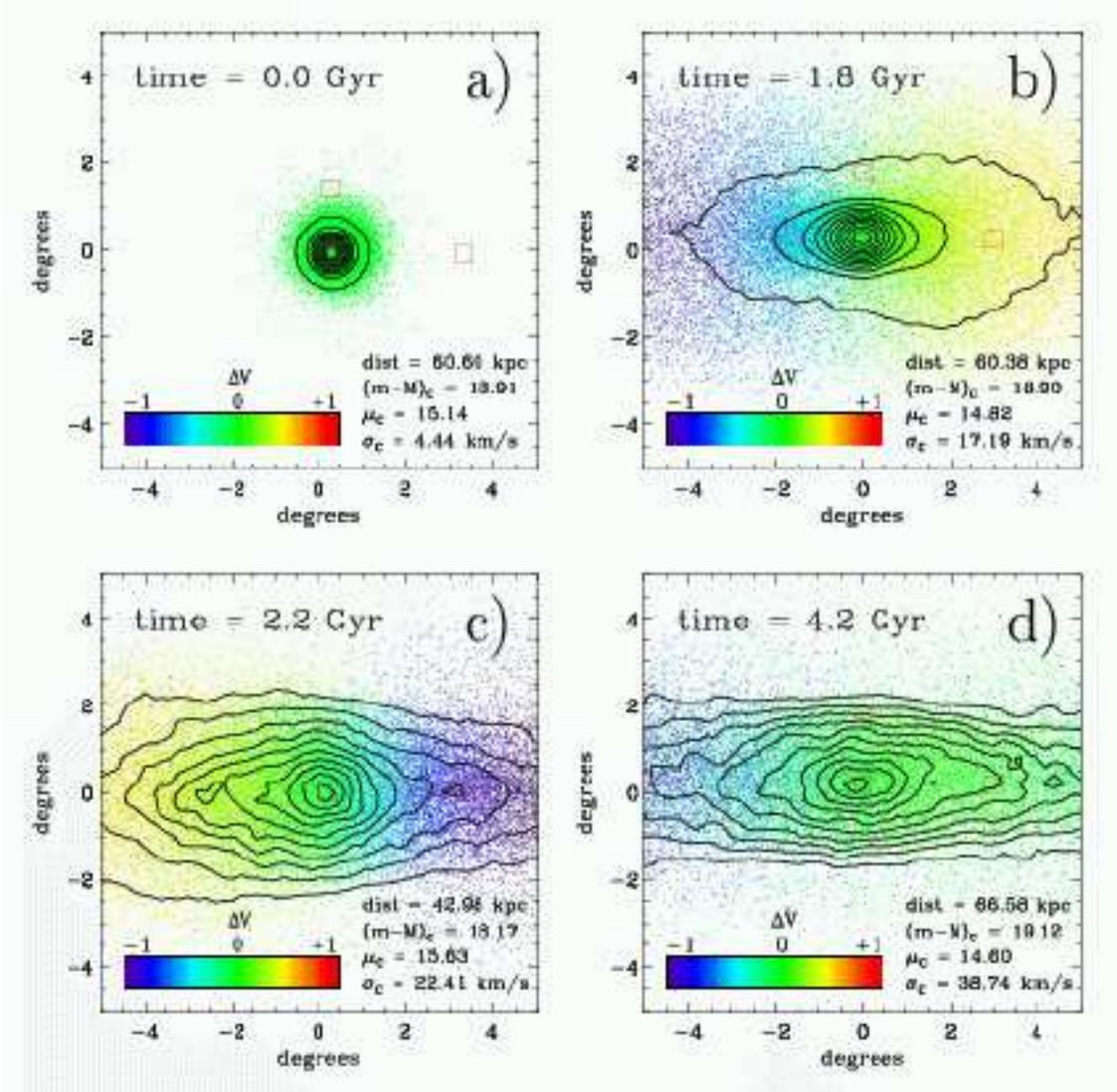}}
\end{picture}
\end{center}
\caption{\label{fig:contours}The model dwarf galaxy {\em Sat-M1} as an
observer on Earth would see it on the sky at four different times of
its evolution.  The first image (a) shows the initial equilibrium
Plummer sphere at a distance of $60.6\,$pc. In (b), the satellite has
just gone through the second apogalacticon, and (c) it has passed
through the third perigalacticon and approaches again an
apogalacticon. Finally, in (d), the dSph galaxy is close to the fifth
apogalacticon. The corresponding evolutionary time is indicated in the
top part of each plot. The contour lines indicate the number density
of stars, with a line spacing equal to 1/11 of the measured dynamic
range of the linear number density. Each star in the image is
color-coded by the shift in $V$-magnitude due to distance modulus
variations relative to the distance modulus $(m-M)_{\rm c}$ of the
center-of-mass. Its distance and the corresponding value $(m-M)_{\rm
c}$ are indicated in the lower part of the plot. Also the observed
central surface brightness $\mu_{\rm c}$ and the central line-of-sight
velocity dispersion $\sigma_{\rm c}$ is listed. Red boxes indicate the
three sub-fields of size $0.5^{\circ}\times 0.5^{\circ}$ for which
CMD's are obtained.  }
\end{figure*}

\section{Horizontal Branch Width}
\label{sec:HB-thick}
Figure \ref{fig:contours} illustrates what a terrestrial observer
would ``see'' on the sky when observing {\em Sat-M1} at four different
stages of its evolution. We depict the satellite when it is close to
apogalacticon, as due to its eccentric orbit, this is where the
satellite spends most of its time and is most likely to be observed.
Each image shows a $10^{\circ}\times 10^{\circ}$ field on the sky
centered on the stellar density maximum.  After the second
perigalactic passage in the tidal field of the Milky Way (i.e.\ after
$t \sig 1.5\times10^9\,$yr) the satellite becomes completely unbound,
but remains visible to the terrestrial observer as diffuse
enhancement of stellar density for several billion years.  This
remnant looks remarkably similar to the dSph galaxies observed today.
If virial equilibrium is assumed, the measured central LOS velocity
dispersion $\sigma_{\rm c}$ implies $M/L$ ratios much larger than the
true value $M/L|_{\rm true} = 3$ (see Figure 6 in KB98).

The satellite remnant in Figure \ref{fig:contours} is quite elongated
and aligned with the orbital trajectory. As the satellite dissolves
and populates extended tidal tails, the density maximum of the observed
dSph traces the trajectory of the original satellite and fades in time
as stars drift apart along the tidal streams.  The elongated
morphological appearence of the remnant for most of its orbit is the
result of this tidal debris seen in projection.  Besides the isopletic
contours of the remnant from star counts, Figure \ref{fig:contours}
shows individual stars in the figure color-coded according to their
distance moduli relative to the center.  There are systematic distance
variations along the semi-major axis, i.e.\ along the orbital path.
The effect is clearly noticeable in the full $10^{\circ}\times
10^{\circ}$ field, i.e. when imaging the entire galaxy beyond several
half light radii. The distance variations are difficult to detect in
small fields.  This demonstrates the necessity for wide field imaging
surveys of the Milky Way satellite galaxies.  To illustrate this,
three sub-fields of $0.5^{\circ}\times 0.5^{\circ}$ are considered as
well, one at the center of the dSph, one off by $3^{\circ}$ along the
orbital path, and the other one off by $1.5^{\circ}$ perpendicular to
that.

\begin{figure*}[ht]
\unitlength1cm
\begin{center}
\begin{picture}(17.0,15.0)
\put( 1.0, 0.0){\epsfbox{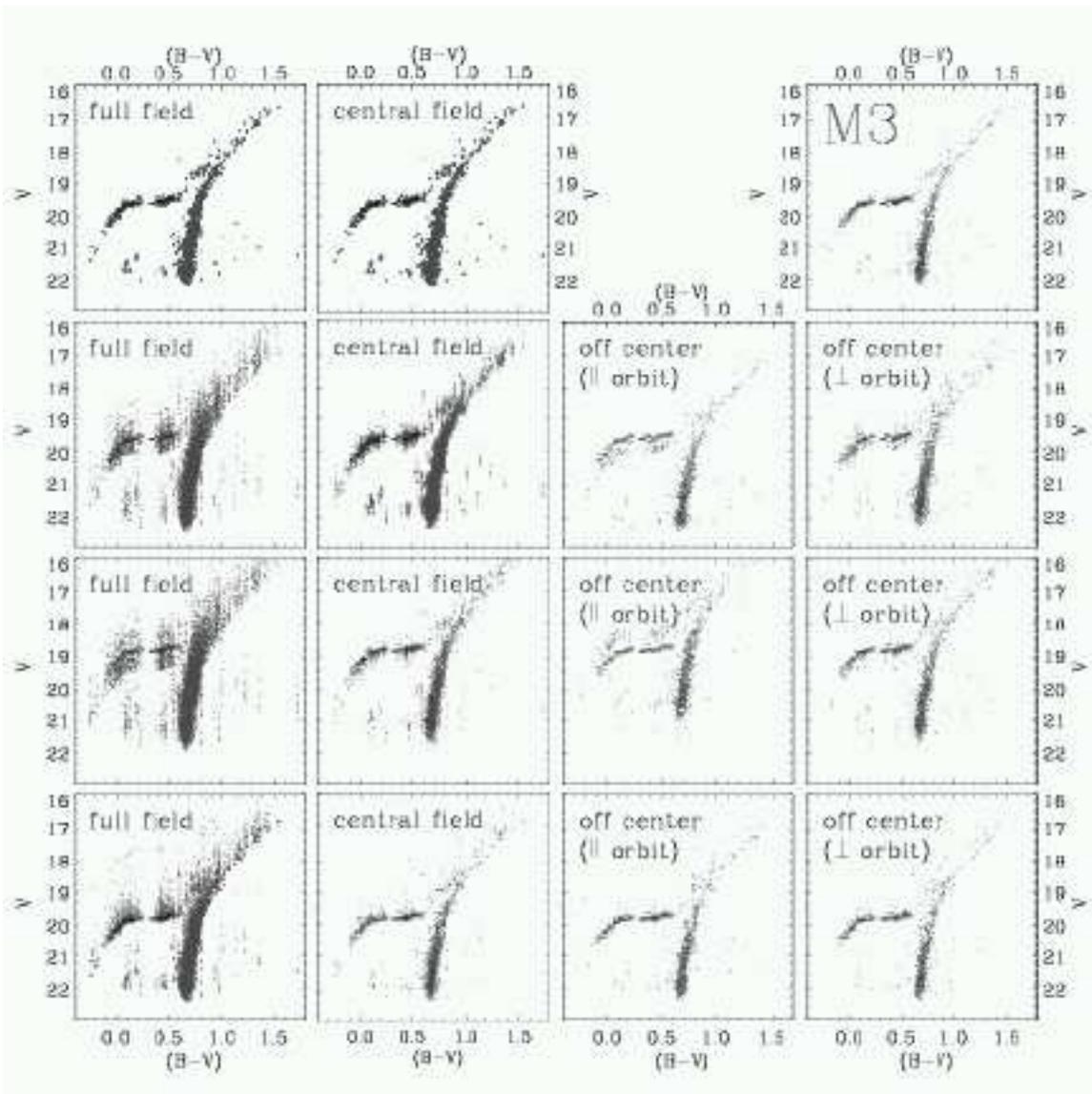}}
\end{picture}
\end{center}
\caption{\label{fig:CMD}Synthetic CMD of the dwarf
  galaxy {\em Sat-M1} as observed on ``Earth'' at four different
  stages of its dynamical evolution. The {\em upper panel} shows the
  initial configuration at $t=0$. The {\em second row} depicts the
  system immediately after the second apogalacticon, i.e.\ at
  $t=1.8\,$Gyr, {\em third row} indicates $t=2.3\,$Gyr shortly before
  the third apogalacticon, and the {\em bottom panel} describes the
  dwarf galaxy at $t=4.0\,$Gyr before its fifth apogalaction. The left
  column gives the CMD of the entire $5^{\circ}\times 5^{\circ}$ field
  shown in Figure \ref{fig:contours}, the second column focuses on
  a $0.5^{\circ}\times 0.5^{\circ}$ field in the center of the dSph
  galaxy, the third column gives the CMD from a $0.5^{\circ}\times
  0.5^{\circ}$ field off-center by $3^{\circ}$ towards the West (along
  the orbital trajectory of the satellite galaxy), and the CMD's in
  the fourth column are obtained from a $0.5^{\circ}\times
  0.5^{\circ}$ field shifted $1.5^{\circ}$ to the North (perpendicular
  to the orbital plane), as indicated by the red boxes in Figure
  \ref{fig:contours}.  For comparison the CMD of the globular cluster M3 is plotted
  in the upper right corner of the figure (the off-center sub-fields
  at $t=0$ contain no stars). To guide your eye for estimating the
  effect of distance variations within the tidal galaxy, the HB position 
  and width of M3 are indicated in the individual CMD's of the  dSph
  (assuming the globular cluster would lie at a  similar distance as
  the  center of density of the dwarf).
 }
\end{figure*}

At this point, a note of caution is necessary. The two model galaxies
discussed here resemble the observed Galactic dSph's in many aspects
(see KK98), however, their surface density profiles are somehow
shallower and more extended than typical observed ones. There are
several reasons for this. (1) To directly build up on the previous
investigation of KK98, we consider two models with apogalactic
distances of only $60\,$kpc. The dwarf galaxies are relatively close
to the hypothetical observer on Earth (who is at a galactic distance
of 8.5kpc) and their projections on his/her plane of the sky appear
therefore relatively large. The Galactic dSph's all have larger
distances and subsequently smaller apparent angular diameter.  (2)
Already in their initial configuration (see \S\ref{sec:model}), the
two model galaxies are more extended than most observed dSph's. During
tidal disruption they only spread out even further. (3) The considered
satellite galaxies both have an initial mass of $10^7\,$M$_{\odot}$.
Given that the remnant only contains a few percent of the total mass,
this starting value may be a bit small compared to most observed
dSph's. (4) The orbital eccentricities of the model galaxies are only
moderately high. However, the most compact morphology in projection is
obtained in the tidal model for viewing angles close to the tangent to
the orbit. As then the long axis of the tidal stream becomes normal to
the plane of the sky, and stars basically ``pile up'' along the line
of sight without strong excursions perpendicular to it. The object has
maximal depth and minimal extent on the sky.  Therefore, dwarf
spheroidal galaxies that are very compact and exhibit a lack of
significant numbers of extra-tidal stars (like Draco, see Odenkirchen
et al.\ 2001, or Aparicio et al.\ 2001) must originate from initial
satellites that move on highly eccentric orbits. This defines an
additional test for the tidal models once reliable proper motions for
the dSph's of the Milky Way become available.  A study aimed at
describing in detail individual Galactic satellites is beyond the
scope of this paper. It requires an extensive (and expensive)
investigation of large areas of the parameter space, and therefore
remains as work for the future.

\begin{figure*}[t]
\unitlength1cm
\begin{center}
\begin{picture}(17.0,8.0)
\put( -3.0,-16.0){\epsfbox{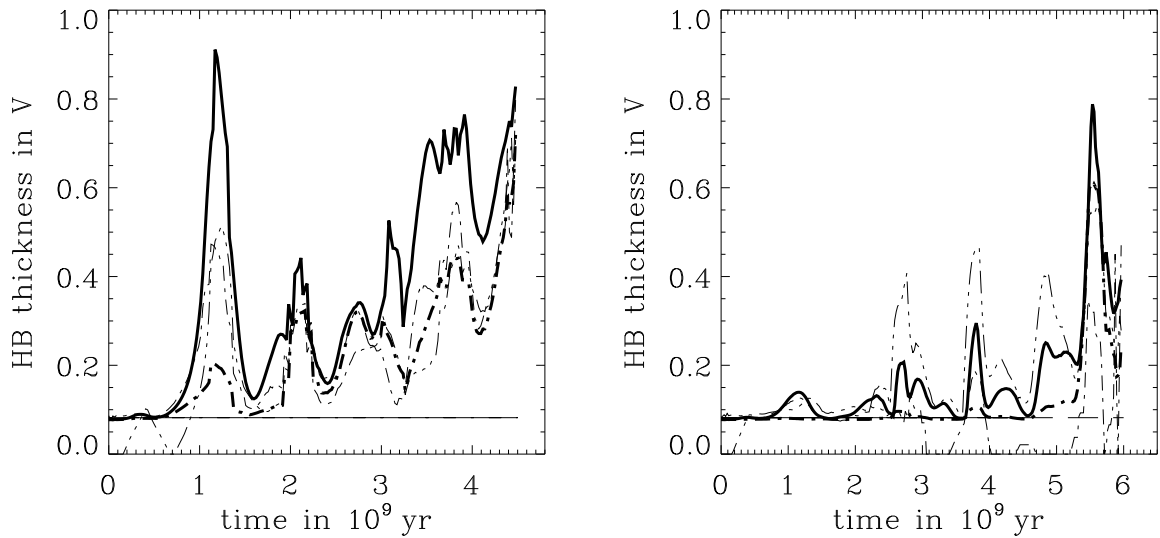}}
\end{picture}
\end{center}
\caption{\label{fig:HB-thickness}Time evolution of the horizontral
branch (HB) thickness in V for models {\em Sat-M1} and {\em
Sat-M2}. The HB thickness is determined as the  average over the  standard
deviations $\sigma(V)$ of stars in 14 color bins of width $\Delta(B-V) =
0.05$ in the color range $-0.1 \le B-V \le 0.6$. The thick full line
indicates the HB width for the CMD of {\em all} stars in the full
$10^{\circ}\times 10^{\circ}$ field, the thick dashed line
takes only stars in the central $0.5^{\circ}\times 0.5^{\circ}$
sub-field into account. The  dashed-dotted and dashed-double-dotted
thin lines indicate thickness measurements from the two sub-field along and
perpendicular to the orbital trajectory of the galaxy (as indicated by
the red boxes in Figure \ref{fig:contours}). For comparison, the
long-dashed thin line indicates the HB thickness of the initial
satellite $\sigma(V)=0.08$. The dotted black line shows the result for
 {\em all} stars on the sky associated with  the tidal stream of the
disrupted satellite (which may be  determined by kinematic selection).
}  
\end{figure*}

The resulting CMD's for the full model galaxy {\em Sat-M1} and for the three
sub-fields are shown in Figure \ref{fig:CMD}.  To clearly detect any
HB thickening in the CMD due to distance effects one needs to consider
the wide-field image.  Only immediately after the complete disruption of the
satellite, i.e.\ when the stars have not had enough time to disperse
significantly, is the stellar density  still high enough to yield
a statistically significant result in the central field. At later
times and for the off-center sub-fields this is not the case, the HB
does not appear significantly widened. It is, however, noticeable that
the sub-fields which are displaced along the major axis (i.e.\ along
the orbital trajectory) exhibit HB's that, although not widened, are
systematically shifted to higher or lower apparent magnitudes relative
to the central sub-field, depending on whether they trace the nearby
or the faraway part of the tidal stream.  With accurate photometry,
this opens the possibility to still test the tidal scenario, even
without a wide-field survey, by taking small fields along the major
axis of the satellite and comparing the mean apparent magnitudes of the
HB stars.  Any systematic variation with position along the axis may
indicate distance moduli variations due to the tidal effect. However,
field contamination by Galactic foreground stars may pose a possible
observational problem.  Especially the red part of the HB appears in a
region of the CMD that may be populated by halo turnoff stars. The
blue part of the HB is less susceptible to this effect, but even
there, some field stars are to be expected.

The time dependence of the HB thickness for both satellites is shown
in Figure \ref{fig:HB-thickness}. After the perigalactic passage that
completely dissolves each satellite, i.e. after $t \sig
1.5\times10^9\,$yr for {\em Sat-M1} and $t \sig 5\times10^9\,$yr in
case of {\em Sat-M2} (see KK98 for details), the HB of the satellite
remnant becomes permanently and significantly widened compared to the
bound stage. The HB width in the full field tends to be about 1.5 to 2
times larger than in the individual sub-fields. If dwarf galaxies are
required to move on orbits with eccentricities larger than considered
here in order to reach better morphological agreement with the
observations, then the estimates for the expected HB width derived in
this paper have to be seen as lower limits.

\begin{figure*}[t]
\unitlength1cm
\begin{center}
\begin{picture}(17.0,15.0)
\put( 0.0, 0.0){\epsfbox{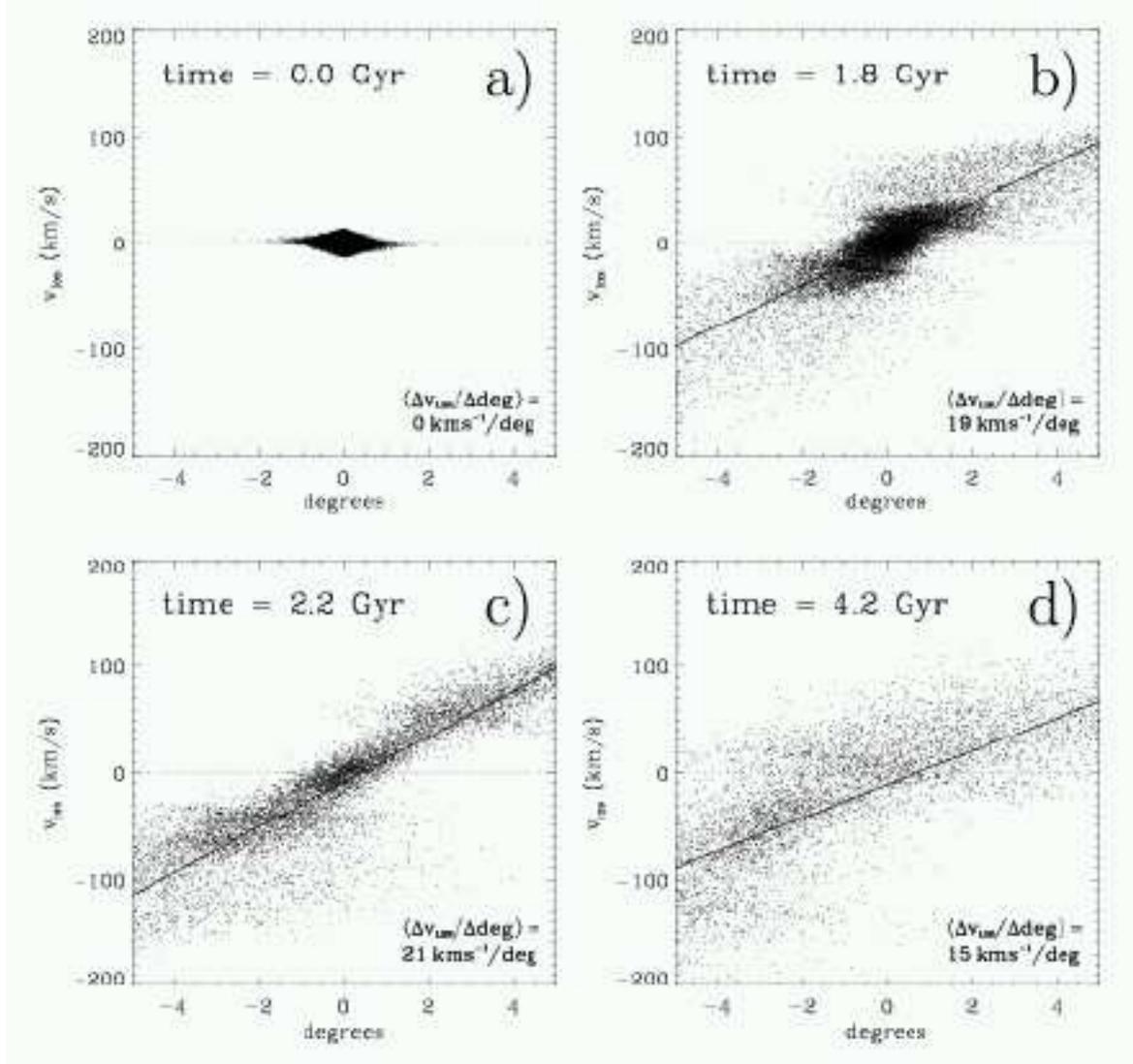}}
\end{picture}
\end{center}
\caption{\label{fig:velocity}Line-of-sight (LOS) velocity gradient for
the model dwarf galaxy {\em Sat-M1} at four different times of its
evolution. The LOS velocity is plotted versus position for stars
selected in a slice of width $0.5^{\circ}$ along the major axis of the
galaxy. The selected times are identical to Figure \ref{fig:contours}.
The tidal model predicts velocity gradients along the orbital
trajectory, for typical projections this corresponds to the major axis
of the galaxy. The expected LOS velocity gradients are of the order of
$20\,$km$\,$s$^{-1}/$deg.}
\end{figure*}

\section{Radial Velocity}
\label{sec:LOS-velocity}
The relation between distance modulus and position along the
semi-major axis of the dSph galaxy (Figure \ref{fig:contours})
introduces a similar correlation between the line-of-sight (LOS)
velocity and position, as illustrated in Figure \ref{fig:velocity}.
Projecting the tidal debris of the disrupted satellite into the
observer's sky leads to a significant LOS gradient along the orbital
trajectory, simply because for larger distances from the stellar
density maximum (along the tidal stream and subsequently on the plane
of the sky) stars lag behind or advance ahead faster. This can be
determined in a spectroscopic survey along the major axis of the dSph
galaxy, as for typical projections the orbital path corresponds to the
major axis of the galaxy. Only for extremely small viewing angles with
respect to the tangent to the trajectory it is possible for a
terrestrial observer to obtain projections of the orbital path onto
the minor axis. This, however, requires very high orbital
eccentricities and appears unlikely.  The spectroscopic probe of the
tidal scenario again requires covering large areas in order to be able
to determine the LOS velocity gradient sufficiently well. The
expected LOS velocity gradients are typically of the order of
$20\,$km$\,$s$^{-1}$ per degree (see also \S4.2 in K97).

\section{Discussion and Summary}
\label{sec:discussion}
Dwarf galaxies are excellent systems for testing current theories of,
and alternatives to, dark matter.  If they are dark matter dominated,
their large mass-to-light ratios make the velocity dispersion profile
or the rotation curve a clean measure of the dynamical mass in dark
matter at all radii.  Recent observations of gas-rich dwarfs show a
nearly solid-body rotation curve, which implies a finite core inside a
few hundred pc.  This is inconsistent with cold dark matter (CDM)
models (van den Bosch et al.\ 2000).  Numerical simulations of CDM
models invariably produces a diverging, power-law ``cusp'' of dark
matter density $\rho \propto r^{-\gamma}$ with $\gamma=1.25 \pm 0.25$
at the scale of a few hundred pc (Navarro, Frenk \& White 1996, Moore
et al., 1998). This failure of CDM also cannot be rescued by
violent feedback from star formation and supernovae
(Gnedin \& Zhao 2001).  The CDM models are furthermore problematic because
they predict the formation of too many dwarf satellites for a
Milky-Way-sized galaxy.  These problems are some of the driving
motivations to look for alternatives to CDM.  Many people have taken
this as a need to assign new properties to the dark matter
(self-interacting dark matter, warm dark matter etc.), or to modify
the Newtonian law of gravity.  Here we reexamined the non-equilibrium
models of dwarf spheroidal galaxies proposed earlier (K97, KK98).
These models are the most conservative models which introduces neither
non-Newtonian dynamics nor any dark matter.  In the non-equilibrium
models, the observed dSph galaxies are identified as the unbound
remnants of tidally disrupted satellite galaxies on an eccentric orbit
with two tidal arms extending along a small angle from the line of
sight of a terrestrial observer.  We demonstrated that these models
predict a detectable spread in distance modulus and a radial velocity
gradient during most of the evolution after tidal disintegration.

The depth effect could be detected with wide-field photometric imaging
of the Galactic dSph galaxies.  To achieve statistical significance it
is necessary to survey a dwarf galaxy out to several half-mass
radii. This is feasible with the recent wide-field cameras (e.g.\ at
the 2.2m telescope of ESO at La Silla).  Distance modulus variations
are best studied with horizontal branch stars outside the instability
strip as they are bright (and easily detectable for the Galactic
dSph's) and have very similar luminosity over a wide range of colors
and metallicities. If the Galactic dSph's are of tidal origin, then
the width of the horizontal branch due to the distance effect is
expected to exceed its intrinsic width significantly. The models
furthermore predict a systematic variation of distance moduli with
position across the galaxy. This can be used to discriminate the
effect from HB widening due to stellar evolution and the star
formation history, because these should not correlate with position.

The non-equilibrium models furthermore predict a strong line-of-sight
velocity gradient along the orbital trajectory, hence radial velocity
data of the Galactic dSph is needed not only to test the tidal
scenario, but also to establish alternative dark matter models and
ultimately the dark matter profiles in dwarfs.

Kleyna et al. (1998) for example tried to measure the distance
gradient effect between two central fields within 30 arcmins of Ursa
Minor.  They found the projection angle is between $90^\circ$ (no
gradient) to $17^\circ.5$.  Unfortuately this is only a $1\sigma$
result, and it would be interesting to extend the study to outer
fields where the effect would be more prominent.  New data coming from
Draco (Aparicio, Carrera, \& Mart{\'\i}nez-Delgado 2001) and Ursa
Minor (Mart{\'\i}nez-Delgado et al.\ 2001 -- which confirms the
double-centerness of Ursa Minor) begin to set the first constraints on the
present model.   
No obvious variation of distance modulus is seen in 
high quality multiband INT WFC images of the central 
$40\arcmin \times 40\arcmin$ of Ursa Minor and Draco
(M. Irwin 2001, private communication).  
There is an indication of velocity gradient over a similar sized
field in the recently completed spectroscopic data set of Draco stars,
with a slope of about 12 km$\,$s$^{-1}$/deg, but curiously orthogonal to the 
expected orbital path of Draco (J. Kleyna 2001, private communication,
see however Figure 4a in K97).
To further constrain the tidal models it may be ultimately necessary
to carry out the difficult task of finding stars
outside the main body of the dwarfs, and following up with spectroscopy.
A certain number of  extra-tidal stars is expected for these dwarfs, given the
strong tidal forces they experience in the Milky Way halo.

We have reported results from two tidal models with quite typical
physical parameters.  It was the aim of this paper to illustrate the
expected effects and suggest possible observational tests rather than
to describe one of the observed dSph's in particular.  The overall
parameter space for the tidal models (e.g.\ satellite mass and density
profile, orbital eccentricity and apogalactic distance) is very
large. Although our general conclusions will remain valid,
quantitative estimates of the HB thickness as well as direction and
strength of a possible velocity gradient depend on the adopted initial
conditions. Therefore applying the proposed test to Galactic dwarfs
requires some readjustments, for example shifting of the off-center control
fields (typically they can be placed closer to the center of surface
density as the observed dwarfs are somewhat less extended than the
projections in our models).  A more comprehensive quantitative
analysis of the tidal model requires the statistical analysis of a
large number of model calculations. This remains work for the future.

How will our result change if some amount of dark matter is present in
the dwarf spheroids?  Without further simulations, we suspect that our
results remain qualitatively the same. Our present models may be
generalized to included cold dark matter if we assume it follows the
same profile as the stars.
 Indead, a small dark halo may not be
enough to bind the system together.  This is supported observationally
by the nearly complete disruptions in the Sagittarius galaxy at 16 kpc
and the SMC at 60 kpc.  The CDM models predict a virial radius of
roughly $10\,$kpc $\left({\sigma / 10\,{\rm kms}^{-1}}\right)$ for a
pristine NFW halo in the field, which is much bigger than the observed
tidal truncation radius of the dwarfs in the Milky Way halo.  So it is
likely that the tidal shocks can significantly modify the density
structure of a dwarf galaxy, both the stellar density profile and the
profile of any dark matter halo.

\acknowledgements{We thank Josef Fried, Eva Grebel,
Mike Irwin, Jan Kleyna, Pavel Kroupa, Doug Lin, Bryan Miller, \& Mark
Wilkinson for stimulating discussions. And we grateful to the comments
and suggestions by our referee Tad Pryor whose careful reading helped to improve and
clarify the arguments presented here.  RSK
acknowledges support by a Otto-Hahn-Stipendium of the
Max-Planck-Gesell\-schaft and subsidies from a NASA astrophysics
theory program supporting the joint Center for Star Formation Studies
at NASA-Ames Research Center, UC Berkeley, and UC Santa
Cruz.
}


\begin{references}
%
\reference{Aparicio} Aparicio, A., Carrera, R., \&
Mart{\'\i}nez-Delgado, D. 2001, \aj, submitted
%
\reference{BDD01} Bekki, K., Couch, W.\ J., Drinkwater, M.\ J. 2001,
\apj, 552, L105 
%
\reference{bounanno93} Bounanno, R., Corsi, C.\ E., Buzzoni, A.,
Cacciari, C., Ferraro, F.\ R., Fusi Pecci, F. 1994, \aap, 290, 69
%
\reference{deblock01} de Blok, W.\ J.\ W., McGaugh, S.\ S., Bosma, A.,
\& Rubin, V.\ C. 2001, \apj, 552, L23
%
\reference{ebi}  Ebisuzaki, T., Makino, J., Fukushige, T., Taiji, M., Sugimoto, D.,
  Ito, T., Okumura, S. 1993, \pasj, 45, 269 
%
\reference{gz01} Gnedin, O., Zhao, H., 2001, \mnras,  submitted (astro-ph/0108108)?
%
\reference{grebel97} Grebel E.\ K. 1997, Reviews in Modern Astronomy, 10, 29
%
\reference{HW99} Helmi, A., White, S.\ D.\ M. 1999, \mnras, 307, 495
%
\reference{IGI94} Ibata, R.\ a., Girmore, G., \& Irwin, M.\ J. 1994,
\nat, 370, 194
%
\reference{irwin95} Irwin, M., Hatzidimitriou, D. 1995, \mnras, 277, 1354
%
\reference{JSH99} Johnston, K.\ V., Sigurdsson, S., Hernquist, L.
1999, \mnras, 302, 771
%
\reference{Ka98} Kleyna, J. Geller, M. Kenyon S., Kurtz, M., Thorstensen J. 1998, \aj, 115, 2359
%
\reference{KK98} Klessen, R.\ S., Kroupa, P. 1998, \apj, 498, 143
%
\reference{K97} Kroupa, P. 1997, New Aston., 2, 139
%
\reference{K93} Kuhn, J.\ R. 1993, \apj, 409, L13
%
\reference{mateo98} Mateo, M. 1998, \araa, 36, 435
%
\reference{m01a} Mart{\'\i}nez-Delgado, D., Alonso-Garc{\'\i}a, J.,
Aparicio, A., \& G{\'o}mez-Flechoso, M.\ A. 2001a, \apj, in press (astro-ph/0101456)
%
\reference{m01a} Mart{\'\i}nez-Delgado, D., G{\'o}mez-Flechoso, M.\
A., Aparicio, A., \& Alonso-Garc{\'\i}a, J.  2001b, \aj, in
preparation
\reference{mr} Moore B. et al. 1998, \apj, 499, L5
%
\reference{nfw} Navarro J.F., Frenk C.S., \& White S.D.M., 1996, \apj, 462, 563
%
\reference{Oh95} Oh, K.\ S., Lin, D.\ N.\ C., Aarseth, S.\ J. 1995, \apj, 442, 142
%
\reference{PP95} Piatek, S., Pryor, C. 1995, \aj, 109, 1071
%
\reference{sandage53} Sandage, A.\ R. 1953, \aj, 58, 61
%
\reference{sandage82} Sandage, A.\ R., Katem, B. 1982, \aj, 88, 1146
%
\reference{sugimoto}  Sugimoto, D., Chikada, Y., Makino, J., Ito, T., Ebisuzaki, T.,
  Umemura, M. 1990, \nat, 345, 33 
%
\reference{vdbosch01} van den Bosch, F.\ C., Robertson, B.\ E.,
Dalcanton, J.\ J., \& de Blok, W.\ J.\ G. 2000, \aj, 119, 1579
%
\end{references}
\end{document}